\documentclass{jfm}
\usepackage{graphicx}
\usepackage{epstopdf, epsfig}
\usepackage{xcolor}
\definecolor{green}{RGB}{20, 180, 80}
\definecolor{blue}{RGB}{20, 80, 242}
\definecolor{red}{RGB}{190, 20, 42}

\shorttitle{Microlayer formation during heterogeneous bubble nucleation}
\shortauthor{M. Saini, X. Chen, S. Zaleski and D. Fuster}

\title{Direct numerical simulations of  microlayer formation during heterogeneous bubble nucleation}

\author{M. Saini\aff{1}
  \corresp{\email{mandeep.saini@sorbonne-universite.fr}},
  X. Chen\aff{1},
  S. Zaleski \aff{1,2}  \corresp{\email{stephane.zaleski@sorbonne-universite.fr}}
 \and D. Fuster\aff{1}
 \corresp{\email{daniel.fuster@sorbonne-universite.fr}}}

\affiliation{\aff{1}Sorbonne Universite and CNRS, 
 Institut Jean le Rond $\partial$'Alembert UMR 7190, F75005 Paris, France
\aff{2}Institut universitaire de France, UMR 7190, Institut Jean le Rond $\partial$'Alembert F75005 Paris, France}

\begin{document}

\maketitle

\begin{abstract}
In this article, we present direct numerical simulation results for the expansion of spherical cap bubbles attached to a rigid wall due to a sudden drop in the ambient pressure. The critical pressure drop beyond which the bubble growth becomes unstable is found to match well with the predictions from classical theory of heterogeneous nucleation imposing a quasi-static bubble evolution. When the pressure drop is significantly higher than the critical value, a liquid microlayer appears between the bubble and the wall. In this regime, the interface outside the microlayer grows at an asymptotic velocity that can be predicted from the Rayleigh--Plesset equation, while the contact line evolves with another asymptotic velocity that scales with a visco-capillary velocity that obeys the Cox--Voinov law. 
In general, three distinctive regions can be distinguished: the region very close to the contact line where dynamics is governed by visco-capillary effects, an intermediate region controlled by inertio-viscous effects away from the contact line yet inside the viscous boundary layer, and the region outside the boundary layer dominated by inertial effects.
The microlayer forms in a regime where
the capillary effects are confined in 
a region much smaller than the viscous boundary layer thickness. In this regime, the global capillary number takes values much larger then the critical capillary number for bubble nucleation, and the microlayer height is controlled by viscous effects and not surface tension.
\end{abstract}



\section{Introduction}
The study of bubble nucleation and the factors controlling the process of bubbles formation began with the influential work of \citet{harvey1945,harvey1946}, who hypothesized that small amounts of gas (nuclei) could adhere to hydrophobic surfaces or surface impurities and become unstable when the pressure difference exceeded the Laplace pressure. Thereafter, many researchers \citep{galloway1954experimental,strasberg1959,greenspan1967radiation} have tried to measure the threshold beyond which observable cavitation occurs. A unified view for the theory of heterogeneous bubble nucleation was provided by \citet{apfel1970} and then by \citet{atchley1989} who proposed a theoretical model to predict cavitation thresholds for spherical cap-shaped nuclei adhered in a conical crevice during quasi-static bubble evolution. In relatively recent studies, \citet{borkent2009} experimentally validated the predictions of quasi-static theories for nucleation threshold by etching well-controlled micron-sized pits on silicon wafers, and reducing the pressure using tension waves generated by transducer. \citet{fuster2014} discussed the stability of bubbly liquids from a free energy perspective and extended these ideas to clusters of bubbles. \citet{saini2021direct} and \citet{saini2022direct} discussed the effect of wall boundary condition on the stability of spherical cap shaped nuclei attached to a rigid wall and showed that pinning can stabilize the bubble nuclei. 

The decrease in pressure beyond the nucleation threshold results in an unstable expansion of the small gas nuclei. During the bubble expansion, a liquid layer (few microns thick) can get trapped between the solid and the bubble, referred to as a microlayer. The formation of a microlayer was studied for the laser-generated bubbles near the solid boundaries by \citet{hupfeld2020dynamics}. The microlayer formation is crucial in the boiling heat transfer applications as it significantly enhances the heat transfer rates (see \citet{judd1976comprehensive}). Yet, little is known about the hydrodynamics of microlayer formation at early stages. \citet{mikic1970bubble,lien1969bubble} and \citet{sullivan2022inertio} demonstrated that at short times, and for high Jacob numbers, the growth of the bubbles in boiling is similar to that in cavitation because of strong inertial effects. 
\citet{guion2018simulations} performed numerical simulations of microlayer formation using a two-phase incompressible solver by modelling the bubble expansion with a uniformly distributed source. \citet{urbano2018direct} and \citet{huber2017direct} performed the direct numerical simulations with mass transfer effects for studying the formation of a microlayer. \citet{burevs2022comprehensive} studied the formation of microlayer using a thin-film based sub-grid model for the velocity of the contact line. \citet{pandey2018bubble} studied the bubble nucleation in boiling using a subgrid microlayer model for microlayer. There are several other studies available in the literature, but there is no common consensus for the structure and the growth of the microlayer \citep{sinha2022microlayer,jung2018hydrodynamic,zou2018origin}. In the cavitation process, microlayer formation influences the bubble shape at the maximum volume, which is an important parameter that afterwards governs the collapse dynamics \citep{saini2022dynamics}. 

The process of microlayer formation is similar to the process of dewetting transition and the deposition of a thin liquid film on the solid surface, well known as the Landau--Levich--Derjaguin (LLD) film \citep{landau1988dragging,derjaguin1993thickness}. This has been extensively studied in the past for setups where a plate is suddenly pushed into or pulled out of the liquid \citep{wilson1982drag,eggers2004hydrodynamic,bonn2009wetting}. More recently, there have been various numerical studies for these problems by \citet{afkhami2018transition,kamal2019dynamic,ming2023early} and \citet{zhang2023time}. A clear understanding of the relationship between LLD film deposition and microlayer formation has not been established yet. A particular problem relevant for microlayer formation is the one considered by \citet{bretherton1961motion} who discussed the motion of bubbles in capillary tubes. A striking difference between Bretherton's problem and the microlayer formation is that the capillary numbers in the former case are much smaller then unity whereas they are comparable to unity for the latter case.
\citet{aussillous2000quick} analysed the film deposition for large capillary numbers and showed that the film height can exhibit a non-monotonic behaviour.

In this article, we use direct numerical simulations to gain insights into the process of heterogeneous bubble nucleation and the formation of a microlayer during the bubble growth. Section \ref{sec:setup} gives the details of our numerical method and setup. In Section \ref{sec:threshold} the nucleation thresholds are discussed for transition from stable bubble oscillations regime to the unstable growth regime. In Section \ref{sec:microlayer}, the dynamics of microlayer formation is examined and the transition from no microlayer to microlayer formation regimes is analyzed. In section \ref{sec:shape}, the structure of the microlayer is described and the scaling arguments are presented for the height of the microlayer to compare it with our numerical results.

\section{Problem set-up} \label{sec:setup}
\begin{figure}
    \centering
    \includegraphics{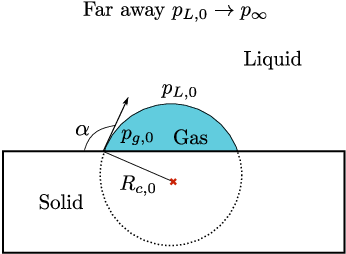}
    \caption{A simplified schematic of the problem.}
    \label{fig:setup}
\end{figure}
We numerically study the response of a spherical cap nucleus to a sudden drop in system pressure using the set-up shown in figure \ref{fig:setup}. We restrict ourselves to an axisymmetric configuration where the initial bubble pressure $p_{g,0}$ is uniform inside the bubble and the liquid, initially at rest, is suddenly exposed to a lower pressure far from the bubble $p_\infty < p_{g,0}$. This problem corresponds to a regime where the bubble growth is predominantly controlled by the inertial effects as the heat and the mass diffusion effects are neglected. Consequently, our study is valid at short times after the bubble nucleation, particularly $t < L_{\mathcal{D}}^2/\mathcal{D}$, where $L_{\mathcal{D}}$ is a characteristic length for diffusion and $\mathcal{D}$ is the Fickian constant for the diffusion law of heat/mass transfer process \citep{bergamasco2017oscillation,sullivan2022inertio}. To render the problem dimensionless, we define the initial bubble radius of curvature as a characteristic length $L_c = R_{c,0}$, the liquid density as a characteristic density $\rho_c = \rho_l$ and the asymptotic Rayleigh--Plesset velocity as a characteristic velocity of the problem $U_c = \sqrt{\frac{2}{3}\frac{\Delta p}{\rho}}$. The dynamics of these bubbles is governed by compressible Navier-Stokes equations which, for a two-phase ($i \in \{l,g\}$) system in a dimensionless form, reads
\begin{eqnarray}
        \frac{\partial \tilde{\rho}_i}{\partial \tilde{t}} + \boldsymbol{\nabla} \boldsymbol{\cdot} (\tilde{\rho}_i \tilde{{\mathbf u}}_i) & = & 0, \label{eq:dnsrho}\\
     \frac{\partial \tilde{\rho}_i \tilde{{\mathbf u}}_i}{\partial \tilde{t}} + \boldsymbol{\nabla} \boldsymbol{\cdot} (\tilde{\rho}_i \tilde{{\mathbf u}}_i \tilde{{\mathbf u}}_i) + \boldsymbol{\nabla} \tilde{p}_i & = &  \frac{\textrm{Oh}^2}{\textrm{Ca}} \boldsymbol{\nabla} \boldsymbol{\cdot} (2 \tilde{\mu}_i \tilde{\boldsymbol{\mathbf{D}}}_i), \label{eq:dnsmom}\\
     \frac{\partial (\tilde{\rho}_i \tilde{E}_i)}{\partial \tilde{t}} + \boldsymbol{\nabla} \boldsymbol{\cdot} (\tilde{\rho}_i \tilde{E}_i \tilde{\mathbf{u}}_i) + \boldsymbol{\nabla} \boldsymbol{\cdot} (\tilde{{\mathbf u}}_i \tilde{p}_i) & = &  \frac{\textrm{Oh}^2}{\textrm{Ca}} \boldsymbol{\nabla} \boldsymbol{\cdot} (2 \tilde{\mu}_i \tilde{\boldsymbol{\mathbf{D}}}_i) \tilde{{\mathbf u}}_i, \label{eq:dnsenergy}
\end{eqnarray}
\noindent where the subscript `$i$' represents the value of a particular variable for the $i^{th}$ component, $\textrm{Ca} = \mu_l U_c/\sigma$ is the capillary number, $\sigma$ is the surface-tension between the liquid and gas phase, $\mu_l$ is the liquid viscosity, $\textrm{Oh} = \mu_l/\sqrt{\rho_l \sigma R_{c,0}}$ is the Ohnesorge number, $\tilde{\mathbf{u}}$ is the dimensionless velocity vector field, $\tilde{\rho}$ is the dimensionless density, $\tilde{p}$ is the dimensionless pressure field, $\tilde{E}$ is the dimensionless total energy per unit volume which is defined as the sum of the internal energy and the kinetic energy $(\tilde{\rho}_i \tilde{e}_i + \frac{1}{2} \tilde{\rho}_i \tilde{\mathbf{u}}_i \boldsymbol{\cdot} \tilde{\mathbf{u}}_i)$, and $\tilde{\boldsymbol{\mathbf{D}}}_i$
is the strain rate tensor.\\

The system of equations is closed by an equation of state for each component. We use a stiffened gas equation of state (EOS) similar to that of \citet{cocchi1996treatment}, given as
\begin{equation}
    \tilde{\rho}_i \tilde{e}_i = \frac{\tilde{p}_i + \tilde{\Gamma}_i \tilde{\Pi}_i}{\tilde{\Gamma}_i - 1},
    \label{eq:EOS}
\end{equation}
where $\tilde{\Gamma_i}$ and $\tilde{\Pi}_i$ are empirical constants taken from \citet{johnsen2006implementation} to replicate the speeds of sound in water and ideal gas.
The interface between the two immiscible fluids is mathematically represented with the Heaviside function $(\mathcal{H})$ which takes the value 1 in the reference component and 0 in the non-reference component \citep{tryggvason2011}. The evolution of the interface is described by an advection equation
\begin{equation}
    \frac{\partial \mathcal{H}}{\partial \tilde{t}} + \tilde{\mathbf{u}} \boldsymbol{\cdot \nabla} \mathcal{H} = 0,
    \label{eq:heavyside}
\end{equation}
\noindent where $\tilde{\mathbf{u}}$ is the average local velocity of the interface which is imposed to be equal to local fluid velocity. The interface conditions are required to couple the motion of fluids in each component. In the absence of mass transfer effects, these conditions are as follows: the velocity is continuous across the interface such that $[[\tilde{\mathbf{u}}]] = 0$, where $[[\cdot]]$ represents the jump in the particular quantity across the interface. The pressure in both the components is related by the Laplace equation 
\begin{equation}
    \frac{1}{\textrm{Oh}^2}[[\tilde{p}]] = - \frac{1}{\textrm{Ca}^2} \tilde{\kappa} + \frac{1}{\textrm{Ca}}[[ \mathbf{n}_I \cdot 2 \tilde{\mu} \tilde{\mathbf{D}} \cdot \mathbf{n}_I ]],
    \label{eq:laplacejump}
\end{equation}
\noindent where $\tilde{\kappa}$ is the dimensionless curvature of the interface and $\mathbf{n}_I$ is the unit vector normal to the interface. We also assume that there is no heat transfer across the interface, so the normal derivative of internal energy remains continuous across the interface i.e. $[[\partial \tilde{e}/\partial n]] = 0.$\\

We integrate (\ref{eq:dnsrho})-(\ref{eq:dnsenergy}) and (\ref{eq:heavyside}) (satisfying Equation (\ref{eq:EOS}) and (\ref{eq:laplacejump})) on the finite volume grids using the numerical method discussed by \citet{fuster2018} also used by \citet{fan2020optimal,saini2022direct,saini2022dynamics,sainiprf} and extended to include thermal effects by \citet{saade2023multigrid}. This method is implemented in the free software program called Basilisk \citep{popinet2015quadtree}. In this implementation, a geometric volume of fluid (VoF) method with piece-wise linear constructions in the discrete cells is used for the interface representation \citep{tryggvason2011}. In the VoF method, the phase characteristic function is represented with the colour function $C_i$ which is $1$ in the reference phase, $0$ in non-reference phase and in the cells containing both phases, $C_i$ takes fractional value between 0 and 1. The conserved quantities (density $C_i \tilde{\rho}_i$, momentum $C_i \tilde{\rho}_i \tilde{\mathbf{u}}_i$, total energy $C_i \tilde{\rho}_i \tilde{E}_i$) are advected consistently with the colour function (see \citep{arrufat2021mass}).\\ 


The capillary forces are added as continuum surface forces (CSF) where the delta function is approximated as the gradient of colour function $\vert \boldsymbol{\nabla} C \vert$ (\cite{brackbill1992continuum,popinet2009accurate,popinet2018numerical}). The primitive variables are density, momentum, total energy and the colour function leading to a density-based formulation where an auxiliary equation is solved to obtain a provisional value of the volume-averaged pressure $\tilde{p}_{\rm{avg}}$ required to reconstruct the pressure of the individual phases required to compute the fluxes of the primitive quantities \citep{kwatra2009method}

\begin{equation}
    {\frac{1}{ \tilde{\rho}_{\rm{avg}} \tilde{c_e}^{2}}} \left( \frac{\partial \tilde{p}_{\rm{avg}}}{\partial \tilde{t}} + \tilde{\mathbf{u}} \boldsymbol{\cdot} \boldsymbol{\nabla} \tilde{p}_{\rm{avg}} \right) 
 = \boldsymbol{\nabla \cdot} \tilde{\mathbf{u}}, \label{eq:prav}
\end{equation}

\noindent where the subindex \emph{``avg''} stands for a volume-averaged quantity and $\tilde{c}_e$ is the effective speed of sound of the mixture. Note that generalized expressions of the equation above accounting for full viscous and thermal effects can be found from \cite{saade2023multigrid} and \citet{urbano2022semi}. For more details about the details of the specific numerical method used here, the reader is referred to the work of \cite{fuster2018}.\\

Similar to the work of \citet{afkhami2018transition,guion2018simulations,kamal2019dynamic} and \citet{ming2023early}, the motion of the contact line is regularized by the Navier-slip model 

\begin{equation}
    \tilde{\mathbf{u}}^{\mathbf{T}} = \frac{\lambda_{num}}{R_{c,0}} \frac{\partial \tilde{\mathbf{u}}^{\mathbf{T}}}{\partial \mathbf{n}_w}    
\end{equation}
\noindent where $\tilde{\mathbf{u}}^{\mathbf{T}}$ is the tangential velocity vector at the wall, $\mathbf{n}_w$ is unit vector normal to the wall and $\lambda_{num}$ is the slip length. In this article, the slip length is fixed to $\lambda_{num} = 0.01 R_{c,0}$ unless mentioned otherwise. This choice is limited by the minimum grid size which is $\Delta_{min} =  0.003 R_{c,0}$ in our case. Some results especially in the region very close to contact line can depend on this choice, however, the results in the microlayer region are independent (see the supplementary materials). We also use a static contact angle model implemented using the approach of \citet{afkhami2008height}. For other boundaries, we have used reflective boundary conditions: $\tilde{\mathbf{u}}^{\mathbf{n}} = 0$, ${\partial \tilde{\mathbf{u}}^{\mathbf{T}}}/{\partial \mathbf{n}} = 0$,
$\mathbf{n} \boldsymbol{\cdot} \boldsymbol{\nabla} \tilde{p} = 0$, where $\tilde{\mathbf{u}}^{\mathbf{n}}$ is the velocity vector normal to the wall and $\mathbf{n}$ is the unit vector normal to the wall. The numerical domain is made large enough ($100 R_{c,0} \times 100 R_{c,0}$) and the numerical grid is coarsened near the boundary to prevent any reflected waves from influencing the bubble dynamics.

\section{Nucleation threshold \label{sec:threshold}}

\begin{figure}
    \centering
    \includegraphics[scale = 0.8]{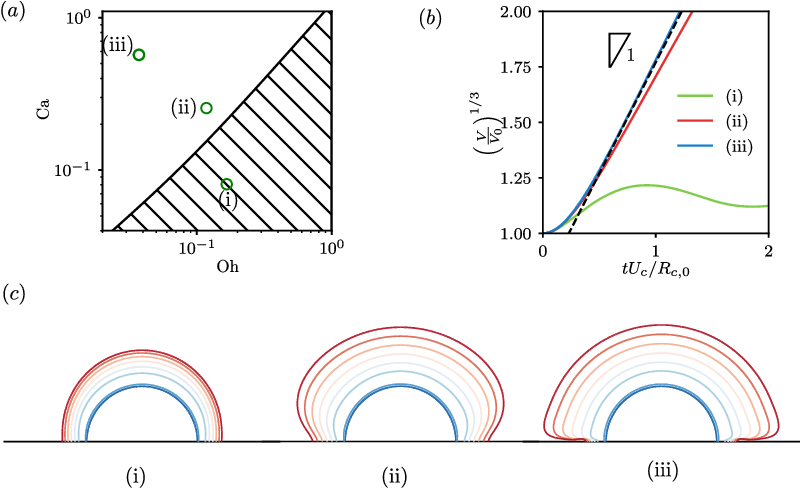}
    \caption{$(a)$ Stability diagram of the cavitation bubbles in the $\textrm{Oh}-\textrm{Ca}$ plane (\ref{eq:cac}) and the location of three representative points: $(\textrm{i})-(\textrm{Oh} = 0.17,\textrm{Ca} = 0.08)$; $(\textrm{ii}) - (\textrm{Oh} = 0.11,\textrm{Ca} = 0.26)$ and $(\textrm{iii})$ - $(\textrm{Oh} = 0.03,\textrm{Ca} = 0.57)$ in the stable bubble regime, unstable bubble without microlayer and unstable bubble with microlayer regime, respectively. $(b)$ The evolution of the equivalent radius for three representative cases as obtained from DNS. The dotted line (with slope 1) is predicted from Rayleigh--Plesset model with negligible acceleration. $(c)$ The evolution of bubble shapes is shown for three representative cases where the colour map corresponds to different dimensionless times. For each case, $tU_c/R_{c,0} \in \{0,0.2,0.5,0.7,0.8,0.9,1.2,1.3\}$.}
    \label{fig:incptcaoh}
\end{figure} 

The nucleation threshold for the growth of a cavitation bubble from a small nuclei (air pocket) is the critical pressure beyond which this nuclei becomes unstable and grows explosively. This critical pressure can be predicted by quasi-static theory for bubble expansion, as shown by \citet{atchley1989,crum1979tensile,fuster2014} and many others. In the case of a hemispherical cap bubble attached to a flat wall, the critical pressure ($p_{cr}$) is
\begin{equation}
     \frac{p_{cr}}{p_{L,0}} = - \left(\frac{3}{2} \frac{p_{L,0}R_{c,0}}{\sigma} \gamma \left(1 + 2 \frac{\sigma}{p_{L,0}R_{c,0}} \right) \right)^{\frac{1}{1 - 3\gamma}} 2 \frac{\sigma}{p_{L,0}R_{c,0}} \left(1 - \frac{1}{3 \gamma}\right).
     \label{eq:pcr}
\end{equation}

\noindent where $p_{L,0}$ is the initial liquid pressure just outside the interface and $\gamma = 1.4$ is the ratio of specific heats for air. For deriving this relation, we have assumed that the bubble contact angle remains 90 degrees and that the contact line can freely move on the solid boundary. In a previous study (see \citet{saini2021direct} and \citet{saini2022direct}), we have shown that the change in contact angle and contact line motion imposes secondary effects on the nucleation threshold. Equation (\ref{eq:pcr}) can be converted to the $\textrm{Ca-Oh}$ plane where it imposes a region of stability such that for given $\textrm{Oh}$ (function of bubble size), there is a critical capillary number (function of critical pressure) beyond which a nuclei will become unstable. This critical capillary number is
\begin{equation}
       \textrm{Ca}_c = \textrm{Oh} \sqrt{2 \left(1 - \frac{1}{3\gamma}\right) \left[\frac{3}{2} \left(\frac{\textrm{Ca}_0}{\textrm{Oh}}\right)^2 \gamma \left(1 + 2 \left(\frac{\textrm{Oh}}{\textrm{Ca}_0}\right)^2\right)\right]^{1/(1 - 3 \gamma)}},
    \label{eq:cac} 
\end{equation}
\noindent where $\textrm{Ca}_0$ is defined as $\textrm{Ca}_0 = \frac{\mu_l}{\sigma} \sqrt{p_{L,0}/\rho_l}$. In figure \ref{fig:incptcaoh}$(a)$, we show the stable region predicted from (\ref{eq:cac}) ($\textrm{Ca} \leq \textrm{Ca}_c$) by shading it with hatched lines. The nucleation threshold predicted from the quasi-static theory (\ref{eq:cac}) is also verified with the direct numerical simulations. To discuss the stability of nuclei, we limit ourselves to show two representative cases: a simulation in the stable region  $(\textrm{i}; \textrm{Oh} = 0.17,\textrm{Ca} = 0.08)$ and another one in the unstable regime  $(\textrm{ii}; \textrm{Oh} = 0.11,\textrm{Ca} = 0.26)$. For a detailed discussion on bubble stability using the current numerical method, the reader is referred to section 3.1 of \citet{saini2022direct}. The evolution of the bubble equivalent radius in figure \ref{fig:incptcaoh}$(b)$ shows that in the former case, the bubble reaches a new equilibrium position, while in the latter case, the bubble becomes unstable and the equivalent radius grows linearly in dimensionless time.

It is also interesting to note that in the unstable case $(\textrm{ii})$, the bubble shape also starts to deviate from a spherical cap  as the viscous stresses close to the wall start to become relevant (see figure \ref{fig:incptcaoh}$(b)$). Moving further away from the stability line, either by decreasing $\textrm{Oh}$ or by increasing $\textrm{Ca}$, the viscous stresses become increasingly important in comparison to the surface-tension stress at the interface. Consequently, in case $(\textrm{iii}; \textrm{Oh} = 0.03,\textrm{Ca} = 0.57)$, where for given $\textrm{Oh}$ and $\textrm{Ca}/\textrm{Ca}_c \gg 1$, the liquid layer is trapped in the gap between the bubble and the solid wall forming what is known as a microlayer. Outside the microlayer, the bubble dynamics is primarily governed by the inertial effects and the growth rate of the interface matches well with the asymptotic velocity ($U_c = \sqrt{\frac{2}{3} \frac{\Delta p}{\rho}}$) predicted from the Rayleigh--Plesset model with negligible acceleration. This is consistent with the experimental results of \citet{bremond2006interaction} for the expansion of bubbles from the cylindrical pits and the growing in contact with a rigid wall.

\begin{figure}
    \centering    \includegraphics{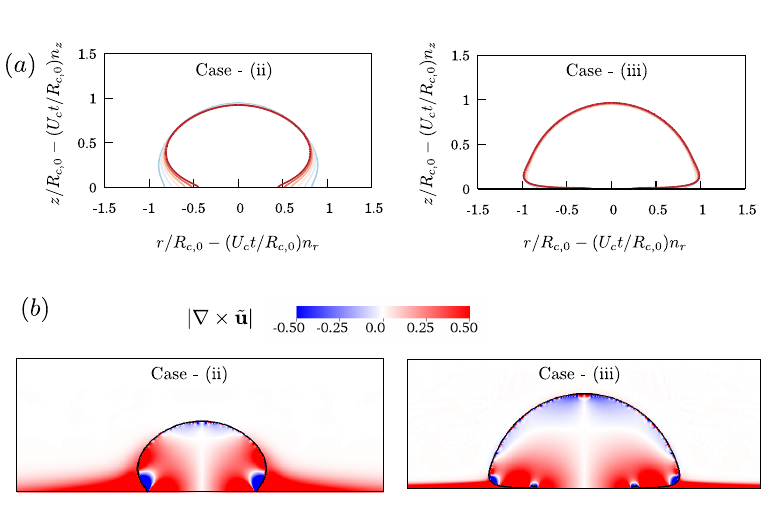}
    \caption{$(a)$ The bubble shapes re-scaled with asymptotic velocity $U_c$ for different instances of times from cases (ii) and (iii) of figure \ref{fig:incptcaoh}$a$. For each case, $tU_c/R_{c,0} \in \{0,0.2,0.5,0.7,0.8,0.9,1.2,1.3\}$. $(b)$ Vorticity near the wall during the bubble nucleation for cases (ii) and (iii) of figure \ref{fig:incptcaoh}$a$}
    \label{fig:facets-rescale}
\end{figure}

In figure \ref{fig:facets-rescale}, we show the bubble shapes at different instants re-scaled with the asymptotic velocity $U_c$ for cases $(\textrm{ii})$ and $(\textrm{iii})$ of figure \ref{fig:incptcaoh}. Certainly, the interface velocity outside the boundary layer tends to $U_c$ and the interface shape from different times overlap well when re-scaled with this velocity. However, inside the boundary layer, the bubble shape is harder to describe as a consequence of three competing effects, namely viscosity, surface-tension and inertia depending upon the control parameters ($\textrm{Ca}, \textrm{Oh}$ and $\alpha$). We also visualize the boundary layer in cases $(ii)$ and $(iii)$ by plotting the vorticity field in figure \ref{fig:facetsca}$(b)$. In case $(\textrm{ii})$, Reynolds number $\textrm{Re} = \textrm{Ca}/\textrm{Oh}^2 \approx 20$ is small as compared with the case $(\textrm{iii})$ where $\textrm{Re} \approx 600$, making the boundarylayer thicker in case $(\textrm{ii})$ as compared to case $(\textrm{iii})$. 
This effect can also be seen in figure \ref{fig:incptcaoh}$(b)$ where the slope of the equivalent radius deviates noticeably from unity.

\section{Microlayer formation dynamics \label{sec:microlayer}}
We study now the dynamics of microlayer formation as function of the capillary number $\textrm{Ca}$, Ohnesorge number $\textrm{Oh}$ and the equilibrium contact angle $\alpha$. We identify three characteristic points on the bubble interface whose evolution is used to describe the transition into the microlayer formation regime. These points are shown in figure \ref{fig:defqtyml} and defined as follows. (a) The point where the bubble interface meets the $z$ axis, which is a measure of the bubble height $h(t)$ (the velocity of this point is denoted as $u_h$). (b) The point where the bubble interface meets the $r$ axis, which gives the length of contact at the wall $c(t)$ and characterizes the motion of contact line (the velocity of this point is denoted as $u_{CL}$). (c) The third point is defined as the interface point at the maximum radial distance $(c_m(t))$ from the $z$ axis, which characterizes the width of the bubble (the velocity of this point is denoted as $u_{c_m}$). 

\subsection{Transition to microlayer formation regime}

\begin{figure}
    \centering
    \includegraphics{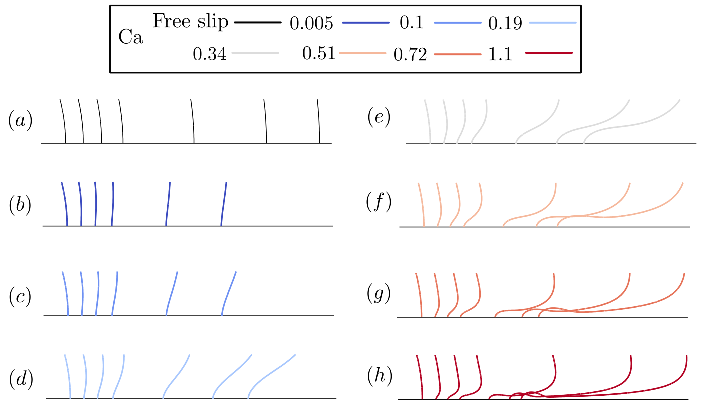}
    \caption{Zoomed-in view of interface shapes near the wall during the bubble expansion for $\alpha = 90^\circ$, $\textrm{Oh} = 0.037$ and varying $\textrm{Ca}$. Each panel corresponds to different capillary numbers $\textrm{Ca}$ represented with the colourmap. In each panel, the interface evolves from left to right and each line corresponds to different non-dimensional time. The contours in each panel are shown at \linebreak $ tU_c/R_{c,0} \in \{0,0.41,0.82,1.22,1.63,2.04,2.44\}$.}
    \label{fig:facetsca}
\end{figure}

We begin by investigating the effect of varying the capillary number $\textrm{Ca}$ for a fixed Ohnesorge number $\textrm{Oh} = 0.037$, fixed value of equilibrium contact angle $\alpha = 90^\circ$ and sliplength $\lambda_{num} = 0.02 R_{c,0}$. The evolution of the velocity of three characteristic points (non-dimensionalized with the characteristic velocity $U_c$) are plotted in figure \ref{fig:Velat3pts}(a-c), for different capillary numbers (colourmap). 
The point $h(t)$ always lies outside the boundary layer for all values of the capillary number and reaches the characteristic velocity $U_c$ after an initial acceleration phase that lasts for the time scales of the order of convective time $(t \approx R_{c,0}/U_c)$. As shown in figure \ref{fig:Velat3pts}$(c)$, for sufficiently large capillary numbers ($\textrm{Ca} > 0.5$), the bubble width velocity $u_{c_m}$  always reaches a constant value which is equal to $U_c$,  while for $\textrm{Ca} < 0.5$, this value  slightly decreases until reaching an asymptotic limit 
around $u_{c_m} \approx 0.8 U_c$ due to the presence of the viscous boundary layer. Remarkably, in the low-capillary-number limit, the contact line velocity $u_{CL}$ also tends to approximately the same value (figure \ref{fig:Velat3pts}$(b)$). In these cases, the interface does not bend much inside the boundary layer as surface-tension stresses dominate over the viscous stresses (see figure \ref{fig:facetsca} $a-d$). 
For high capillary numbers (e.g. $\textrm{Ca} > 0.5$), the viscous stresses become dominant and the contact line velocity at sufficiently large times takes values smaller than $U_c$. 
In this regime, we observe a clear microlayer formation (figure \ref{fig:facetsca} $f-h$). 


\begin{figure}
    \centering
    \includegraphics{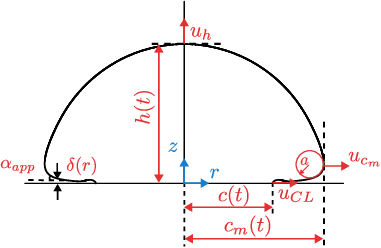}
    \caption{The definition of characteristic points on the bubble interface. These are used to describe the bubble expansion and formation of a microlayer. These points are the height of bubble $h(t)$, the width of bubble $c_m(t)$ and the contact line location $c(t)$.}
    \label{fig:defqtyml}
\end{figure}

\begin{figure}
    \centering
    \includegraphics{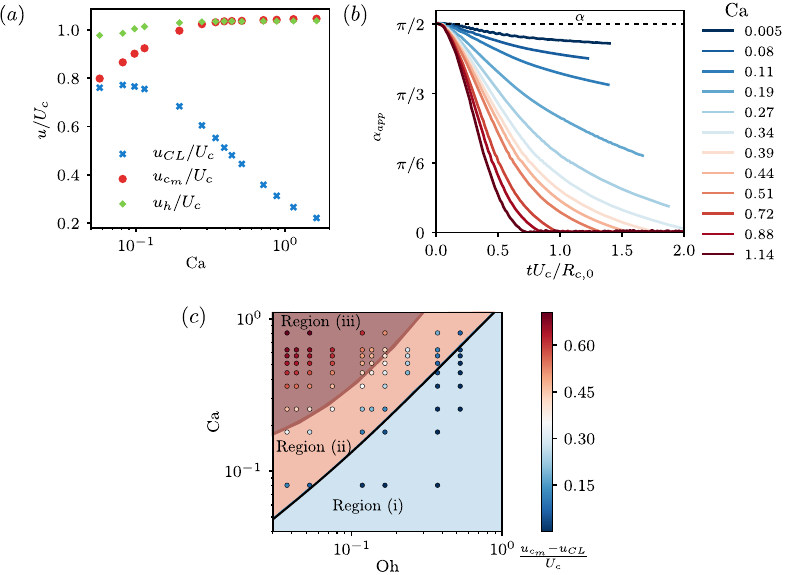}
    \caption{$(a)$ The non-dimensional velocity of three interface points defined in figure \ref{fig:defqtyml} at non-dimensional time $tU_c/R_c = 1.22$ is shown for different values of capillary numbers (for $\alpha = 90^\circ$,$\textrm{Oh} = 0.037$). $(b)$ The minimum value of apparent angle defined as the angle between the tangent to the interface and the radial axis $r$ inside liquid for $\alpha = 90^\circ$, $\textrm{Oh} = 0.037$ and varying $\textrm{Ca}$. $(c)$ The difference in velocity of bubble width and the contact line velocity is plotted with colourmaps in the $\textrm{Ca}-\textrm{Oh}$ plane to characterize three regions. Region $(\textrm{i})$, where bubble remains stable and does not grow; Region $(\textrm{ii})$, where the bubble becomes unstable but the microlayer does not forms and Region $(\textrm{iii})$, where the bubble is unstable and the microlayer forms. The blue line separating regions $(\textrm{i})$ and $(\textrm{ii})$ is the theoretical prediction from (\ref{eq:pcr}) and the beige line separating regions $(\textrm{ii})$ and $(\textrm{iii})$ is the iso-countour of $u_{c_m} - u_{CL} \approx 0.5 U_c$.}
    \label{fig:alphamin}
\end{figure}

The difference between bubble width velocity and the contact line velocity $u_{c_m} - u_{CL}$ is an indicator of whether a microlayer forms or not.
In figure \ref{fig:alphamin}$(a)$, we show the velocity of the three points $h,c,c_m$ in the same plane at an instant $t U_c/R_{c,0} = 1.22$ (after the initial transient). The transition from the bending to a clear microlayer formation regime happens at $\textrm{Ca} \approx 0.5$ in this case.
Because, for $\textrm{Ca} > 0.5$, the velocity of bubble width $u_{c_m}$ is approximately equal to $U_c$, the formation of a microlayer can be discussed solely in terms of the dimensionless contact line velocity $u_{CL}/U_c$. This velocity decreases monotonically with $\textrm{Ca}$ and consequently the growth rate of microlayer length increases monotonically with $\textrm{Ca}$. In section \ref{sec:contactlinevel}, we will analyse the contact line velocity in detail.

\begin{figure}
    \centering
    \includegraphics[scale = 0.78]{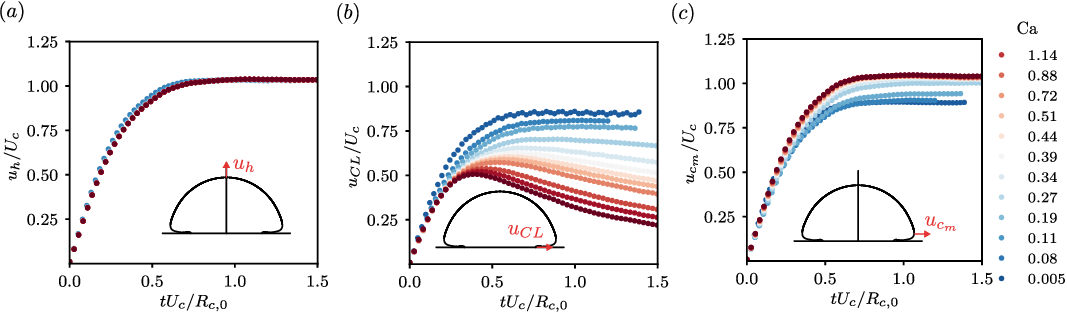}
    \caption{Results for bubble expansion in the case of $\alpha = 90^\circ$,$\textrm{Oh} = 0.037$ and varying $\textrm{Ca}$. This figure characterizes the motion of the interface using three characteristic points defined in figure \ref{fig:defqtyml}. $(a)$ The dimensionless velocity of bubble height ${u_h}/{U_c}$. $(b)$ The dimensionless velocity of contact line ${u_{CL}}/{U_c}$. $(c)$ The dimensionless velocity of bubble width ${u_{c_m}}/{U_c}$}
    \label{fig:Velat3pts}
\end{figure}

To characterize the bending of the interface, we introduce an apparent contact angle $\alpha_{app}$ defined as the minimum angle between the tangent to the interface and the axis parallel to the wall inside a small region near the wall arbitrarily chosen as $z \le 0.25 R_{c,0}$. In this region, the interface height is denoted as $\delta(r)$ (see figure \ref{fig:defqtyml}). For a hemispherical cap, the apparent angle $\alpha_{app}$ is $\pi/2$ at $t=0$ but the interface bending near the contact line can result in a decrease of the minimum apparent contact angle for $t>0$. In figure \ref{fig:alphamin}$(b)$, we plot the temporal evolution of the apparent contact angle $\alpha_{app}$. For small capillary numbers $\textrm{Ca}$, the apparent angle decreases slightly below the $\alpha$. As $\textrm{Ca}$ increases, the interface bends significantly and this angle decreases quickly to zero as the interface becomes parallel to the wall.

The regimes where a microlayer form in the $\textrm{Oh}$ - $\textrm{Ca}$ plane is represented in figure \ref{fig:alphamin}$(c)$ using as colour scale the velocity difference between bubble width and the contact line $u_{c_m} - u_{CL}$. A clear microlayer appears in the cases where $u_{c_m} - U_{CL} \gtrsim 0.5 U_c$ (region \textrm{iii}). Notably, the capillary numbers above which the microlayer forms (shown by the beige curve in figure \ref{fig:alphamin}$c$) depend upon the pressure forcing as compared to the critical pressure predicted for the nucleation (blue curve in figure \ref{fig:alphamin}$c$), a microlayer becoming visible in the cases where $\textrm{Ca}/\textrm{Ca}_c(\textrm{Oh}) \gg 1$. This correlation between the transition into microlayer formation and the critical capillary number for the bubble growth has often been overlooked in previous microlayer formation studies. In these cases, the Reynolds number $\textrm{Re} = \textrm{Ca}/\textrm{Oh}^2 \gg 1$ and the Weber number $\textrm{We} \gg 1$, indicating that the liquid inertia is important for the formation of a microlayer. If the capillary number is comparable to the critical capillary number, the formation of a microlayer is suppressed (region \textrm{ii} in figure \ref{fig:alphamin}$c$).

\begin{figure}
    \centering
    \includegraphics[scale = 0.75]{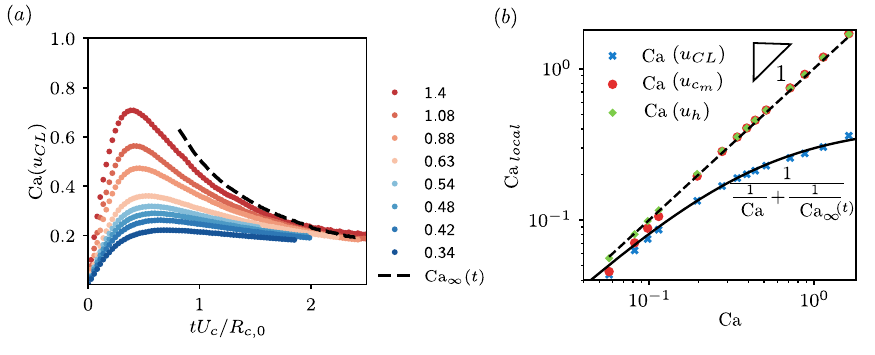}
    \caption{Results for bubble expansion in the case of $\alpha = 90^\circ$,$\textrm{Oh} = 0.037$ and varying $\textrm{Ca}$. $(a)$ The evolution contact line capillary number $\textrm{Ca}(u_{CL}) = \mu u_{CL}/\sigma$ for different values of global capillary numbers $\textrm{Ca}$ (colourmap). Asymptotic value $\textrm{Ca}_\infty(t)$ obtained from fitting the (\ref{eq:hm}) is also shown with a dashed line. $(b)$ Local capillary numbers for the three interface points defined in figure \ref{fig:defqtyml} at non-dimensional time $tU_c/R_c = 1.22$ are shown for different values of global capillary numbers. The dotted line is $\textrm{Ca} = \textrm{Ca}_{local}$ and the solid line is fitted using harmonic averaging (\ref{eq:hm}).}
    \label{fig:uandcaallca}
\end{figure}
\subsection{\label{sec:contactlinevel}Contact line velocity}
Close to the contact line, the flow is mainly governed by visco-capillary effects, thus the velocity scale $\sigma/\mu$ is expected to be a relevant parameter. 
In figure \ref{fig:uandcaallca}, we plot the evolution of the local dimensionless contact line capillary number, $\textrm{Ca}(u_{CL}) = \mu u_{CL}/\sigma$. 
Interestingly, for $\textrm{Ca} > 0.3$ and at $t U_{c}/R_{c,0} > 1$, the contact line capillary number seems to converge to an asymptotic value. We highlight this fact in figure \ref{fig:uandcaallca}$(b)$ (blue cross), where we show that  $\textrm{Ca}_{CL}$ at time $tU_c/R_{c,0} \approx 1.25$  reaches a plateau for large $\textrm{Ca} \to \infty$. A similar asymptotic behaviour was observed in the numerical simulations of \citet{guion2017modeling}. The asymptotic value of the contact line capillary number $\textrm{Ca}_{\infty}(t)$ can be found by fitting the numerical data with the harmonic average formula
\begin{equation}
    \textrm{Ca}(u_{CL}) = \frac{1}{\frac{1}{\textrm{Ca}} + \frac{1}{\textrm{Ca}_{\infty}(t)}},
    \label{eq:hm}
\end{equation}
\noindent where $\textrm{Ca}_\infty (t)$ is the value that the contact line capillary number $\textrm{Ca}(u_{Cl})$ would reach in the limit of $\textrm{Ca}/\textrm{Ca}_c \to \infty$. Equation (\ref{eq:hm}) matches well with the numerical data (see figure \ref{fig:uandcaallca}$b$) obtaining $\textrm{Ca}_\infty (t = 1.25 R_{c,0}/U_c) \approx 0.4$ for $\alpha=90^\circ$. The time evolution of $\textrm{Ca}_\infty(t)$ fitted from \ref{eq:hm} is also shown in figure \ref{fig:uandcaallca}$(a)$ with a dashed line and the in supplementary video 1. We note that the asymptotic contact line capillary number $\textrm{Ca}_\infty(t)$ depends on the contact angle, as discussed next. The local capillary numbers defined with the velocity of bubble height and the velocity of maximum bubble width are equal to the global capillary number $\textrm{Ca}$ (see figure \ref{fig:uandcaallca}$b$) because $U_{c} = u_{h} = u_{c_m}$ (see figure \ref{fig:alphamin}$a$). 

\subsection{\label{sec:alpha}Effect of equilibrium contact angle}

\begin{figure}
    \centering
    \includegraphics[scale = 0.9]{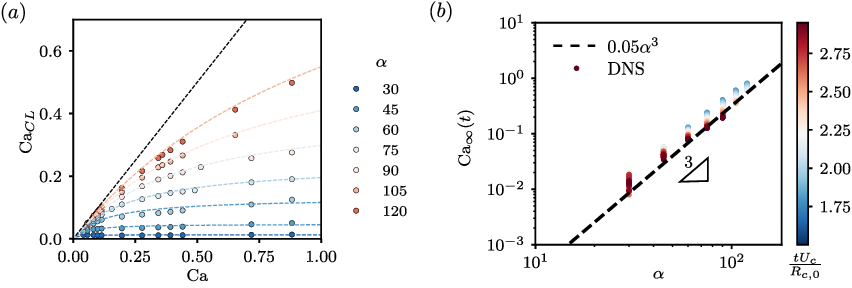}
    \caption{Results for bubble expansion in the case of $\textrm{Oh} = 0.037$ and varying capillary numbers $\textrm{Ca}$ as well as contact angle $\alpha$. $(a)$ The numerical results (circles) for contact line capillary number $\textrm{Ca}(u_{CL})$ as a function of global capillary number $\textrm{Ca}$; the dashed lines show the fitting curves obtained using (\ref{eq:hm}) for different contact angles (colourmap). $(b)$ The evolution of fitting parameter $\textrm{Ca}_\infty (t)$ that is the contact line capillary number for $\textrm{Ca} \to \infty$ is plotted as a function of $\alpha$, along with the (\ref{eq:coxalp0}) (dashed line).}
    \label{fig:alpha}
\end{figure}

We investigate the effect of the equilibrium contact angle $\alpha$ (angle implemented at the smallest grid cell) by performing a parametric study for varying $\alpha \in (30^\circ,135 ^\circ)$ and capillary number $\textrm{Ca} \in (0.07,1.4)$ for fixed Ohnesorge number $\textrm{Oh} = 0.037$. We restrict ourselves to the regime where $\textrm{Ca} > \textrm{Ca}_c$. In figure \ref{fig:alpha}(a), we show the contact line capillary number $\textrm{Ca}(u_{CL})$ with circles as a function of global capillary number $\textrm{Ca}$ for several contact angles $\alpha$ (colourmap), and their respective fitting curves (dotted lines) obtained using \ref{eq:hm}. Small contact angles favour the formation of a microlayer. In particular, for $\alpha < 60^\circ$, the velocity of the contact line remains so small that even for a capillary number slightly above its critical value for nucleation, the microlayer forms almost instantaneously. Similar to the hemispherical nuclei case, for large capillary numbers $\textrm{Ca}$ the contact line capillary number $\textrm{Ca}(u_{CL})$ approaches an asymptotic value $(\textrm{Ca}_\infty)$.  In figure \ref{fig:alpha}$(b)$, we show the variation of $\textrm{Ca}_{\infty}(t)$ obtained by fitting the $\textrm{Ca}(u_{CL})$ with (\ref{eq:hm}) for several times (colourmap). The numerical data for $\textrm{Ca}(u_{CL})$ versus $\textrm{Ca}$ and the fitting with (\ref{eq:hm}) for all times are also shown in the supplementary videos 1 and 2. Remarkably, for all angles, $\textrm{Ca}_{\infty}$ approaches an asymptotic value which is proportional to the cube of the contact angle $\alpha$ (for $\alpha \le 90^\circ$). The cubic relation of the local capillary number with the contact angle can be recovered from the well-known Cox--Voinov law \citep{voinov1976hydrodynamics,cox1986dynamics}, originally derived for an asymptotic limit of small interface slope and small capillary numbers, 
\begin{equation}
    \alpha_{app}^3 = \alpha_m^3 + 9 \frac{\mu u_{CL}}{\sigma} ln\left(\frac{l_o}{l_i}\right),
    \label{eq:cox-voinov}
\end{equation}
\noindent where $\alpha_{app}$ is the apparent contact angle, $\alpha_m$ is the microscopic contact angle at equilibrium, and $l_o/l_i$ is the ratio of outer (macroscopic) length scale and the inner (microscopic) length scale. The contact angle $\alpha$ is the numerical equivalent of $\alpha_m$, as discussed by \citet{afkhami2018transition}. Also, the apparent contact angle $\alpha_{app}$ approaches to zero near the contact line (see figure \ref{fig:alphamin}). Substituting $\alpha_{app} = 0$ and $\alpha_m = \alpha$, we readily obtain a relation between the $\textrm{Ca}(u_{CL})$ and $\alpha$ as

\begin{equation}
    \textrm{Ca}(u_{CL}) = \frac{1}{9 ln(l_o/l_i)} \alpha^{3},
    \label{eq:coxalp0}
\end{equation}

\noindent which gives a cubic dependence of the contact line capillary number $\textrm{Ca}(u_{CL})$ on the contact angle $\alpha$. Furthermore, matching the numerical and theoretical prefactors, i.e. $\frac{1}{9 ln(l_o/l_i)} \approx 0.05$, results in $l_o/l_i \approx 10$. The microscopic length $l_i$ is given by the regularization parameter near the contact line which is the slip length in our set-up that further yields $l_0 = 0.1 R_{c,0}$, which is a reasonable value as it lies between the bounds of the bubble size and the slip length which are the largest and smallest length scales of the problem, that is, $\lambda_{num} < l_0 < R_{c,0}$.


\section{\label{sec:shape}Microlayer morphology}

\begin{figure}
    \centering
    \includegraphics[scale = 0.875]{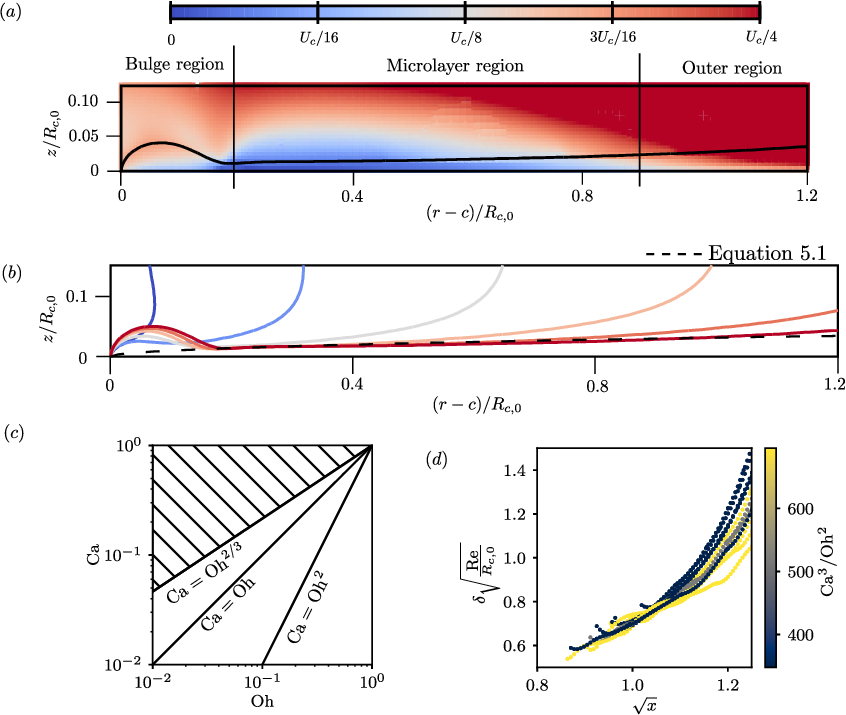}
    \caption{(a) The shape of a microlayer and the velocity magnitude for $\alpha = 90^\circ$, $\textrm{Ca} = 1.4$, $\textrm{Oh} = 0.37$, at $t U_c/R_{c,0} = 2.65$. $(b)$ The evolution of microlayer shape at different dimensionless times $t U_c/R_{c,0} \in \left( 0.45,0.88,1.33,1.77,2.22,2.65 \right)$ for $\alpha = 90^\circ$, $\textrm{Ca} = 1.4$. $(c)$ Different inequalities that indicate the relative importance of viscous, capillary and the inertial effects based on the Ohnesorge and capillary numbers. The hatched region is where the diffusive \citep{cooper1969microlayer} scaling is expected. It approximately coincides with the microlayer formation region in figure \ref{fig:alphamin}$(c)$.
    $(d)$ Microlayer shape rescaled using (\ref{eq:boundarylayer}) for all the cases from figure \ref{fig:alphamin}$(c)$ choosing the constant $C_1 = 1$.}
    \label{fig:mlflow}
\end{figure}

Finally, we characterize the shape of the microlayer in the regimes where the capillary number is much larger than the critical capillary numbers $\textrm{Ca}/\textrm{Ca}_c \gg 1$. Two very distinctive features of the interface near the wall are the formation of a bulge or rim near the contact line, and a long and flat microlayer region (see figure \ref{fig:mlflow}$a$). Similar features were observed by \citet{guion2017modeling} and \citet{urbano2018direct}. figure \ref{fig:mlflow}$a$ and \ref{fig:mlflow}$b$ show that inside the microlayer region, the morphology of the microlayer does not evolve significantly in time as the velocity in this region becomes very small. However, the contact line still moves with a faster velocity due to the surface tension stresses which results in accumulation of liquid in the bulge. \citet{ming2023early} and \citet{guion2017modeling} have discussed the growth of the bulge. In this article, we shift our attention to the characterization of a microlayer region from figure \ref{fig:mlflow}$(a)$. 

\citet{cooper1969microlayer} and \citet{guion2018simulations} predict a square root of time $\sqrt{t}$ behaviour for the growth of the microlayer height from the scaling arguments based on the boundary layer theory 
\begin{equation}
  \delta_{1}(x) \simeq C_1 \sqrt{\frac{\nu x}{U_c}} \equiv C_1 \sqrt{x\frac{R_{c,0}}{\textrm{Re}}},
  \label{eq:boundarylayer}
\end{equation}
where $C_1$ is a constant of the order unity, $x$ is the radial coordinate shifted by the initial location of the contact line $x = r - c_0$ and $\nu$ is the kinematic viscosity in the liquid.
Alternatively, for the Bretherton problem, we know that
the height of a thin liquid film in the limit of $\textrm{Ca} \ll 1$ is 
\begin{equation}
  \delta_{2}(x) \simeq C_2 r_2 \textrm{Ca}^{2/3},
  \label{eq:bretherton}
\end{equation}
where $C_2$ is a constant of the order unity and $r_2$ is the radius of the capillary tube \cite{bretherton1961motion} (in the LLD problems, the capillary length replaces the radius $r_c$, see \citet{landau1988dragging,gennes2004capillarity}).
The characteristic length equivalent to $r_2$ in the microlayer problem is the nose radius $a$ schematically shown in figure \ref{fig:defqtyml} which
varies in time and is related to the relative importance of inertial, viscous and surface-tension effects near the bubble nose.
The largest length scale of the problem is the bubble radius that evolves like $R(t) \sim U_c t$.
The viscous and surface-tension (capillary) effects may also impose smaller length scales.
Viscosity gives the diffusive scale $l_\nu = \sqrt{\nu R(t)/ U_c}$, while the capillary scale is $l_\sigma = \sigma/(\rho U_c^2)$.
The nose radius cannot be larger than the bubble radius, so provided that
\begin {equation}
   R(t) > l, \label{Rbig}
\end{equation}
we have $a=l$ where $l$ is either the diffusive or the capillary scale. 
This smaller length is decided by the relative importance of the viscous and surface-tension effects
\begin {equation}
  l =  \textrm{max}(l_\nu,l_\sigma).
\end {equation}
Equation (\ref{Rbig}) is verified if
\begin {equation}
    \left(\frac{l_\nu}{R(t)}\right)^2 = \frac{\textrm{Oh}^2}{\textrm{Ca}} \frac{R_{c,0}}{R(t)} \ll 1,
    \label{eq:lnurt}
\end{equation}
and
\begin {equation}
    \frac{l_\sigma}{R(t)} = \frac{\textrm{Oh}^2}{\textrm{Ca}^2} \left(\frac{R_{c,0}}{R(t)}\right)^2 \ll 1.
    \label{eq:lsigrt}
\end{equation}
Both of the above inequalities are verified in region (iii) of figure \ref{fig:alphamin}$(c)$.
If we compare the viscous and capillary length, the viscous effects dominate if
\begin {equation}
   \frac{\textrm{Ca}^3}{\textrm{Oh}^2} \frac{R}{R_{c,0}} \gg 1.
   \label{eq:lnusig}
\end{equation}
Again in region (iii) of figure \ref{fig:alphamin}$(c)$, this inequality is verified. We thus expect the boundary layer regime to dominate in region (iii) with
the conclusion that the microlayer thickness is given by the expression of \citet{cooper1969microlayer}, that is, (\ref{eq:boundarylayer}).
\citet{aussillous2000quick} showed that for the bubbles moving in tubes at large capillary numbers, it is possible to observe a transition from the regime where the height of the liquid layer is given by (\ref{eq:bretherton}) to a regime where (\ref{eq:boundarylayer}) applies. In figure \ref{fig:mlflow}$(c)$, the inequalities of \ref{eq:lnurt} - \ref{eq:lnusig} are sketched and the region of validity of the boundary layer theory is shaded with hatched lines. The numerical results shown in figure \ref{fig:mlflow}$(d)$ confirm that for different cases in the shaded region, the scaling given by the boundary layer theory works well as the microlayer shapes for these cases overlap well when re-scaled using (\ref{eq:boundarylayer}).

\begin{figure}
    \centering
    \includegraphics{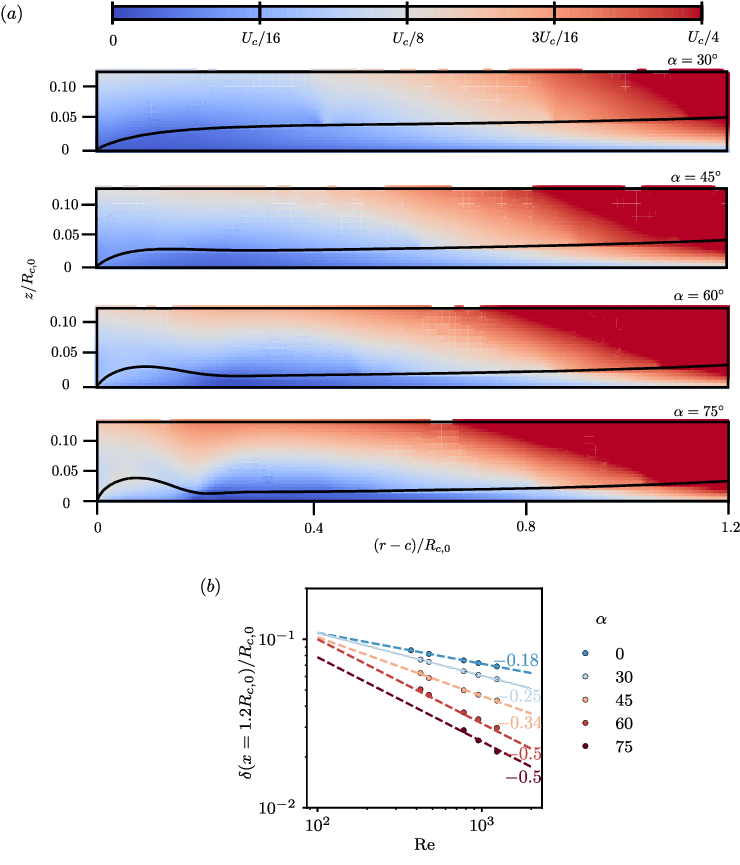}
    \caption{$(a)$ The velocity magnitude and the bubble interface near the wall and inside the microlayer at dimensionless time $t U_c/R_{c,0} = 2.65$, for $\alpha \in (30^\circ, 45^\circ, 60^\circ, 75^\circ)$, $\textrm{Oh} = 0.037$, $\textrm{Ca} = 1.4$. $(b)$ The height of the microlayer at $r = 1.25 R_{c,0}$ as a function of Reynolds number $\textrm{Re}$ for different contact angle $\alpha$ in the cases where a clear microlayer is formed among those described in figure \ref{fig:alpha}$(a)$. Dotted lines indicate power law $(\delta/R_{c,0} = C_3 Re^{m})$ fitting of the numerical data with exponents $m$ written as labels for different angles (colormap).}
    \label{fig:mlheight}
\end{figure}

The effect of the contact angle on the shape of the microlayer shape is shown in figure \ref{fig:mlheight}$(a)$, where we draw the
interface and velocity field near the microlayer at ($t U_c/R_{c,0} = 2.65$) for $\alpha \in \left(30^\circ,45^\circ, 60^\circ,
75^\circ\right)$ and $\textrm{Ca}/\textrm{Ca}_c \gg 1$. The morphology of the interface near the contact line depends on the contact angle,  
the bulge near the contact line disappearing for small contact angles (see figure \ref{fig:mlheight}$a$). As seen from colourmaps in figure \ref{fig:mlheight}$(a)$ and from Section \ref{sec:alpha}, the speed of the contact line decreases markedly for small contact angles; therefore, the liquid accumulation near the contact line and the formation of the bulge are not observed for small angles (roughly $\alpha \lesssim 45^\circ$). The footprint of the bulge also affects the boundary layer scaling given in (\ref{eq:boundarylayer}). Since we are in the regime where $l_\nu \gg l_\sigma$, we expect the microlayer height to scale with the Reynolds number $\textrm{Re}$. {\textcolor{blue} By fitting our numerical data with power laws of the form $C_3 Re^{m}$, where $C_3$ and $m$ are fitting parameters represented in figure \ref{fig:mlheight}$(b)$, we find that $m$ takes a classical value for viscous boundary layer evolution in single phase flows  for large angles (roughly $m=1/2$ for $\alpha \gtrsim 60^\circ$), whereas smaller values of exponent $m$ are predicted for the smaller values of the contact angle.
In these conditions, the interface is shown to significantly influence the classical development
of the viscous boundary layer thickness.}

\section{Conclusions}
We have performed fully resolved direct numerical simulations of the expansion of bubbles from spherical cap nuclei when the pressure far from the bubble changes suddenly. In agreement with the theory of heterogeneous bubble nucleation, the bubble becomes unstable when the capillary number is larger than the critical capillary number for a given Ohnesorge number. A clear microlayer forms in the regimes where the difference between the velocity of the contact line and the bubble width is larger than 0.5 times the inertial velocity scale $u_{c_m} - u_{CL} > 0.5 U_c$. In these cases, the capillary number is much larger than the critical capillary number, while no microlayer is observed when the capillary number is comparable or smaller than this critical capillary number. In the regime where microlayer forms, both Reynolds and Weber numbers are large, signaling the importance of inertial effects. The speed of the apparent contact line in our numerical simulations recover a cubic dependence on the contact angle, as given by the Cox--Voinov law. The capillary effects play a critical role only in the close vicinity of the contact line and are responsible for the formation of the bulge. From scaling arguments, we show that the height of the microlayer is influenced by viscous effects and not by surface tension effects.
We then conclude that a microlayer forms in the regime where the visco-inertial effects are much more important than the capillary effects and propose scaling laws for the evolution of the microlayer height that depend on the Reynolds number and the contact angle.

\backsection[Funding \& Acknowledgements]{This research is supported by European Union (EU) under the MSCA-ITN grant agreement number 813766, under the project named ultrasound cavitation in soft matter (UCOM).}

\backsection[Declaration of interests]{The authors report no conflict of interest.}

\backsection[Author contributions]{D.F. and S.Z. designed the research program. M.S. conducted the simulations and analyzed data and wrote the original draft. M.S., X.C., SZ and D.F.  discussed the results. M.S., S.Z. and D.F. reviewed and edited the article.}
\newpage
\bibliographystyle{jfm}
\bibliography{biblio}

\begin{thebibliography}{59}
\expandafter\ifx\csname natexlab\endcsname\relax\def\natexlab#1{#1}\fi
\def\au#1{#1} \def\ed#1{#1} \def\yr#1{#1}\def\at#1{#1}\def\jt#1{\textit{#1}}
  \def\bt#1{#1}\def\bvol#1{\textbf{#1}} \def\vol#1{#1} \def\pg#1{#1}
  \def\publ#1{#1}\def\arxiv#1{#1}\def\org#1{#1}\def\st#1{\textit{#1}}

\bibitem[Afkhami {\em et~al.\/}(2018)Afkhami, Buongiorno, Guion, Popinet,
  Saade, Scardovelli \& Zaleski]{afkhami2018transition}
{\sc \au{Afkhami, S}, \au{Buongiorno, J}, \au{Guion, A}, \au{Popinet, S},
  \au{Saade, Y}, \au{Scardovelli, R} \& \au{Zaleski, S}} \yr{2018}
  \at{Transition in a numerical model of contact line dynamics and forced
  dewetting}.  \jt{Journal of Computational Physics}  \bvol{374},
  \pg{1061--1093}.

\bibitem[Afkhami \& Bussmann(2008)]{afkhami2008height}
{\sc \au{Afkhami, S} \& \au{Bussmann, M}} \yr{2008}  \at{Height functions for
  applying contact angles to 2d vof simulations}.  \jt{International journal
  for numerical methods in fluids}  \bvol{57}~(4),  \pg{453--472}.

\bibitem[Apfel(1970)]{apfel1970}
{\sc \au{Apfel, RE}} \yr{1970}  \at{The role of impurities in
  cavitation-threshold determination}.  \jt{The Journal of the Acoustical
  Society of America}  \bvol{48}~(5B),  \pg{1179--1186}.

\bibitem[Arrufat {\em et~al.\/}(2021)Arrufat, Crialesi, Fuster, Ling, Malan,
  Pal, Scardovelli, Tryggvason \& Zaleski]{arrufat2021mass}
{\sc \au{Arrufat, T}, \au{Crialesi, M}, \au{Fuster, D}, \au{Ling, Y},
  \au{Malan, L}, \au{Pal, S}, \au{Scardovelli, R}, \au{Tryggvason, G} \&
  \au{Zaleski, S}} \yr{2021}  \at{A mass-momentum consistent, volume-of-fluid
  method for incompressible flow on staggered grids}.  \jt{Computers \& Fluids}
   \bvol{215},  \pg{104785}.

\bibitem[Atchley \& Prosperetti(1989)]{atchley1989}
{\sc \au{Atchley, AA} \& \au{Prosperetti, A}} \yr{1989}  \at{The crevice model
  of bubble nucleation}.  \jt{The Journal of the Acoustical Society of America}
   \bvol{86}~(3),  \pg{1065--1084}.

\bibitem[Aussillous \& Qu{\'e}r{\'e}(2000)]{aussillous2000quick}
{\sc \au{Aussillous, P} \& \au{Qu{\'e}r{\'e}, D}} \yr{2000}  \at{Quick
  deposition of a fluid on the wall of a tube}.  \jt{Physics of fluids}
  \bvol{12}~(10),  \pg{2367--2371}.

\bibitem[Bergamasco \& Fuster(2017)]{bergamasco2017oscillation}
{\sc \au{Bergamasco, L} \& \au{Fuster, Daniel}} \yr{2017}  \at{Oscillation
  regimes of gas/vapor bubbles}.  \jt{International Journal of Heat and Mass
  Transfer}  \bvol{112},  \pg{72--80}.

\bibitem[Bonn {\em et~al.\/}(2009)Bonn, Eggers, Indekeu, Meunier \&
  Rolley]{bonn2009wetting}
{\sc \au{Bonn, D}, \au{Eggers, J}, \au{Indekeu, J}, \au{Meunier, J} \&
  \au{Rolley, E}} \yr{2009}  \at{Wetting and spreading}.  \jt{Reviews of modern
  physics}  \bvol{81}~(2),  \pg{739}.

\bibitem[Borkent {\em et~al.\/}(2009)Borkent, Gekle, Prosperetti \&
  Lohse]{borkent2009}
{\sc \au{Borkent, BM}, \au{Gekle, S}, \au{Prosperetti, A} \& \au{Lohse, D}}
  \yr{2009}  \at{Nucleation threshold and deactivation mechanisms of nanoscopic
  cavitation nuclei}.  \jt{Physics of fluids}  \bvol{21}~(10),  \pg{102003}.

\bibitem[Brackbill {\em et~al.\/}(1992)Brackbill, Kothe \&
  Zemach]{brackbill1992continuum}
{\sc \au{Brackbill, JU}, \au{Kothe, DB} \& \au{Zemach, C}} \yr{1992}  \at{A
  continuum method for modeling surface tension}.  \jt{Journal of Computational
  physics}  \bvol{100}~(2),  \pg{335--354}.

\bibitem[Bremond {\em et~al.\/}(2006)Bremond, Arora, Dammer \&
  Lohse]{bremond2006interaction}
{\sc \au{Bremond, N}, \au{Arora, M}, \au{Dammer, SM} \& \au{Lohse, D}}
  \yr{2006}  \at{Interaction of cavitation bubbles on a wall}.  \jt{Physics of
  fluids}  \bvol{18}~(12),  \pg{121505}.

\bibitem[Bretherton(1961)]{bretherton1961motion}
{\sc \au{Bretherton, FP}} \yr{1961}  \at{The motion of long bubbles in tubes}.
  \jt{Journal of Fluid Mechanics}  \bvol{10}~(2),  \pg{166--188}.

\bibitem[Bure{\v{s}} \& Sato(2022)]{burevs2022comprehensive}
{\sc \au{Bure{\v{s}}, L} \& \au{Sato, Y}} \yr{2022}  \at{Comprehensive
  simulations of boiling with a resolved microlayer: validation and sensitivity
  study}.  \jt{Journal of Fluid Mechanics}  \bvol{933}.

\bibitem[Cocchi {\em et~al.\/}(1996)Cocchi, Saurel \&
  Loraud]{cocchi1996treatment}
{\sc \au{Cocchi, JP}, \au{Saurel, R} \& \au{Loraud, JC}} \yr{1996}
  \at{Treatment of interface problems with {G}odunov-type schemes}.  \jt{Shock
  waves}  \bvol{5}~(6),  \pg{347--357}.

\bibitem[Cooper \& Lloyd(1969)]{cooper1969microlayer}
{\sc \au{Cooper, MG} \& \au{Lloyd, AJP}} \yr{1969}  \at{The microlayer in
  nucleate pool boiling}.  \jt{International Journal of Heat and Mass Transfer}
   \bvol{12}~(8),  \pg{895--913}.

\bibitem[Cox(1986)]{cox1986dynamics}
{\sc \au{Cox, RG}} \yr{1986}  \at{The dynamics of the spreading of liquids on a
  solid surface. {P}art 1. viscous flow}.  \jt{Journal of fluid mechanics}
  \bvol{168},  \pg{169--194}.

\bibitem[Crum(1979)]{crum1979tensile}
{\sc \au{Crum, LA}} \yr{1979}  \at{Tensile strength of water}.  \jt{Nature}
  \bvol{278}~(5700),  \pg{148--149}.

\bibitem[Derjaguin(1993)]{derjaguin1993thickness}
{\sc \au{Derjaguin, BV}} \yr{1993}  \at{On the thickness of a layer of liquid
  remaining on the walls of vessels after their emptying, and the theory of the
  application of photoemulsion after coating on the cine film (presented by
  academician {AN} {F}rumkin on {J}uly 28, 1942)}.  \jt{Progress in Surface
  Science}  \bvol{43}~(1-4),  \pg{129--133}.

\bibitem[Eggers(2004)]{eggers2004hydrodynamic}
{\sc \au{Eggers, J}} \yr{2004}  \at{Hydrodynamic theory of forced dewetting}.
  \jt{Physical Review Letters}  \bvol{93}~(9),  \pg{094502}.

\bibitem[Fan {\em et~al.\/}(2020)Fan, Li \& Fuster]{fan2020optimal}
{\sc \au{Fan, Y}, \au{Li, H} \& \au{Fuster, D}} \yr{2020}  \at{Optimal
  subharmonic emission of stable bubble oscillations in a tube}.  \jt{Physical
  Review E}  \bvol{102}~(1),  \pg{013105}.

\bibitem[Fuster {\em et~al.\/}(2014)Fuster, Pham \& Zaleski]{fuster2014}
{\sc \au{Fuster, D}, \au{Pham, K} \& \au{Zaleski, S}} \yr{2014}  \at{Stability
  of bubbly liquids and its connection to the process of cavitation inception}.
   \jt{Physics of Fluids}  \bvol{26}~(4),  \pg{042002}.

\bibitem[Fuster \& Popinet(2018)]{fuster2018}
{\sc \au{Fuster, D} \& \au{Popinet, S}} \yr{2018}  \at{An all-{M}ach method for
  the simulation of bubble dynamics problems in the presence of surface
  tension}.  \jt{Journal of Computational Physics}  \bvol{374},  \pg{752--768}.

\bibitem[Galloway(1954)]{galloway1954experimental}
{\sc \au{Galloway, WJ}} \yr{1954}  \at{An experimental study of acoustically
  induced cavitation in liquids}.  \jt{The Journal of the Acoustical Society of
  America}  \bvol{26}~(5),  \pg{849--857}.

\bibitem[Gennes {\em et~al.\/}(2004)Gennes, Brochard, Qu{\'e}r{\'e} {\em
  et~al.\/}]{gennes2004capillarity}
{\sc \au{Gennes, PG}, \au{Brochard, F}, \au{Qu{\'e}r{\'e}, D} \& \au{others}}
  \yr{2004} {\em Capillarity and wetting phenomena: drops, bubbles, pearls,
  waves\/}.  \publ{Springer}.

\bibitem[Greenspan \& Tschiegg(1967)]{greenspan1967radiation}
{\sc \au{Greenspan, M} \& \au{Tschiegg, CE}} \yr{1967}  \at{Radiation-induced
  acoustic cavitation; apparatus and some results}.  \jt{J. Res. Natl. Bur.
  Stand., Sect. C}  \bvol{71},  \pg{299}.

\bibitem[Guion(2017)]{guion2017modeling}
{\sc \au{Guion, AN}} \yr{2017}  \at{Modeling and simulation of liquid
  microlayer formation and evaporation in nucleate boiling using computational
  fluid dynamics}. PhD thesis, Massachusetts Institute of Technology.

\bibitem[Guion {\em et~al.\/}(2018)Guion, Afkhami, Zaleski \&
  Buongiorno]{guion2018simulations}
{\sc \au{Guion, A}, \au{Afkhami, S}, \au{Zaleski, S} \& \au{Buongiorno, J}}
  \yr{2018}  \at{Simulations of microlayer formation in nucleate boiling}.
  \jt{International Journal of Heat and Mass Transfer}  \bvol{127},
  \pg{1271--1284}.

\bibitem[Harvey(1945)]{harvey1945}
{\sc \au{Harvey, NE}} \yr{1945}  \at{Decompression sickness and bubble
  formation in blood and tissues}.  \jt{Bulletin of the New York Academy of
  Medicine}  \bvol{21}~(10),  \pg{505}.

\bibitem[Harvey(1946)]{harvey1946}
{\sc \au{Harvey, NE}} \yr{1946}  \at{Decompression sickness and bubble
  formation in blood and tissue, bull}.  \jt{Anesthesiology: The Journal of the
  American Society of Anesthesiologists}  \bvol{7}~(4),  \pg{457--457}.

\bibitem[Huber {\em et~al.\/}(2017)Huber, Tanguy, Sagan \&
  Colin]{huber2017direct}
{\sc \au{Huber, G}, \au{Tanguy, S}, \au{Sagan, M} \& \au{Colin, C}} \yr{2017}
  \at{Direct numerical simulation of nucleate pool boiling at large microscopic
  contact angle and moderate {J}akob number}.  \jt{International Journal of
  Heat and Mass Transfer}  \bvol{113},  \pg{662--682}.

\bibitem[Hupfeld {\em et~al.\/}(2020)Hupfeld, Laurens, Merabia, Barcikowski,
  G{\"o}kce \& Amans]{hupfeld2020dynamics}
{\sc \au{Hupfeld, T}, \au{Laurens, G}, \au{Merabia, S}, \au{Barcikowski, S},
  \au{G{\"o}kce, B} \& \au{Amans, D}} \yr{2020}  \at{Dynamics of laser-induced
  cavitation bubbles at a solid--liquid interface in high viscosity and high
  capillary number regimes}.  \jt{Journal of Applied Physics}  \bvol{127}~(4),
  \pg{044306}.

\bibitem[Johnsen \& Colonius(2006)]{johnsen2006implementation}
{\sc \au{Johnsen, Eric} \& \au{Colonius, Tim}} \yr{2006}  \at{Implementation of
  weno schemes in compressible multicomponent flow problems}.  \jt{Journal of
  Computational Physics}  \bvol{219}~(2),  \pg{715--732}.

\bibitem[Judd \& Hwang(1976)]{judd1976comprehensive}
{\sc \au{Judd, RL} \& \au{Hwang, KS}} \yr{1976}  \at{A comprehensive model for
  nucleate pool boiling heat transfer including microlayer evaporation} .

\bibitem[Jung \& Kim(2018)]{jung2018hydrodynamic}
{\sc \au{Jung, S} \& \au{Kim, H}} \yr{2018}  \at{Hydrodynamic formation of a
  microlayer underneath a boiling bubble}.  \jt{International Journal of Heat
  and Mass Transfer}  \bvol{120},  \pg{1229--1240}.

\bibitem[Kamal {\em et~al.\/}(2019)Kamal, Sprittles, Snoeijer \&
  Eggers]{kamal2019dynamic}
{\sc \au{Kamal, C}, \au{Sprittles, JE}, \au{Snoeijer, JH} \& \au{Eggers, J}}
  \yr{2019}  \at{Dynamic drying transition via free-surface cusps}.
  \jt{Journal of fluid mechanics}  \bvol{858},  \pg{760--786}.

\bibitem[Kwatra {\em et~al.\/}(2009)Kwatra, Su, Gr{\'e}tarsson \&
  Fedkiw]{kwatra2009method}
{\sc \au{Kwatra, Nipun}, \au{Su, Jonathan}, \au{Gr{\'e}tarsson, J{\'o}n~T} \&
  \au{Fedkiw, Ronald}} \yr{2009}  \at{A method for avoiding the acoustic time
  step restriction in compressible flow}.  \jt{Journal of Computational
  Physics}  \bvol{228}~(11),  \pg{4146--4161}.

\bibitem[Landau \& Levich(1988)]{landau1988dragging}
{\sc \au{Landau, L} \& \au{Levich, B}} \yr{1988}  \at{Dragging of a liquid by a
  moving plate}.  \bt{In {\em Dynamics of curved fronts\/}},  \pg{pp.
  141--153}.  \publ{Elsevier}.

\bibitem[Lien(1969)]{lien1969bubble}
{\sc \au{Lien, YC}} \yr{1969}  \at{Bubble growth rates at reduced pressure.}
  PhD thesis, Massachusetts Institute of Technology.

\bibitem[Mikic {\em et~al.\/}(1970)Mikic, Rohsenow \&
  Griffith]{mikic1970bubble}
{\sc \au{Mikic, BB}, \au{Rohsenow, WM} \& \au{Griffith, P}} \yr{1970}  \at{On
  bubble growth rates}.  \jt{International Journal of Heat and Mass Transfer}
  \bvol{13}~(4),  \pg{657--666}.

\bibitem[Ming {\em et~al.\/}(2023)Ming, Qin \& Gao]{ming2023early}
{\sc \au{Ming, H}, \au{Qin, J} \& \au{Gao, P}} \yr{2023}  \at{Early stage of
  bubble spreading in a viscous ambient liquid}.  \jt{Journal of Fluid
  Mechanics}  \bvol{964},  \pg{A41}.

\bibitem[Pandey {\em et~al.\/}(2018)Pandey, Biswas, Dalal \&
  Welch]{pandey2018bubble}
{\sc \au{Pandey, Vinod}, \au{Biswas, Gautam}, \au{Dalal, Amaresh} \& \au{Welch,
  Samuel~WJ}} \yr{2018}  \at{Bubble lifecycle during heterogeneous nucleate
  boiling}.  \jt{Journal of Heat Transfer}  \bvol{140}~(12),  \pg{121503}.

\bibitem[Popinet(2009)]{popinet2009accurate}
{\sc \au{Popinet, S}} \yr{2009}  \at{An accurate adaptive solver for
  surface-tension-driven interfacial flows}.  \jt{Journal of Computational
  Physics}  \bvol{228}~(16),  \pg{5838--5866}.

\bibitem[Popinet(2015)]{popinet2015quadtree}
{\sc \au{Popinet, S}} \yr{2015}  \at{A quadtree-adaptive multigrid solver for
  the {S}erre--{G}reen--{N}aghdi equations}.  \jt{Journal of Computational
  Physics}  \bvol{302},  \pg{336--358}.

\bibitem[Popinet(2018)]{popinet2018numerical}
{\sc \au{Popinet, S}} \yr{2018}  \at{Numerical models of surface tension}.
  \jt{Annual Review of Fluid Mechanics}  \bvol{50},  \pg{49--75}.

\bibitem[Saade {\em et~al.\/}(2023)Saade, Lohse \& Fuster]{saade2023multigrid}
{\sc \au{Saade, Y}, \au{Lohse, D} \& \au{Fuster, D}} \yr{2023}  \at{A multigrid
  solver for the coupled pressure-temperature equations in an all-mach solver
  with vof}.  \jt{Journal of Computational Physics}  \bvol{476},  \pg{111865}.

\bibitem[Saini(2022)]{saini2022direct}
{\sc \au{Saini, M}} \yr{2022}  \at{Direct numerical simulations of nucleation
  and collapse of bubbles attached to walls}. PhD thesis, Sorbonne
  Universit{\'e}.

\bibitem[Saini {\em et~al.\/}(2024)Saini, Saade, Fuster \& Lohse]{sainiprf}
{\sc \au{Saini, Mandeep}, \au{Saade, Youssef}, \au{Fuster, Daniel} \&
  \au{Lohse, Detlef}} \yr{2024}  \at{Finite speed of sound effects on asymmetry
  in multibubble cavitation}.  \jt{(Accepted in Physical Review Fluids)} .

\bibitem[Saini {\em et~al.\/}(2022)Saini, Tanne, Arrigoni, Zaleski \&
  Fuster]{saini2022dynamics}
{\sc \au{Saini, M}, \au{Tanne, E}, \au{Arrigoni, M}, \au{Zaleski, S} \&
  \au{Fuster, D}} \yr{2022}  \at{On the dynamics of a collapsing bubble in
  contact with a rigid wall}.  \jt{Journal of Fluid Mechanics}  \bvol{948},
  \pg{A45}.

\bibitem[Saini {\em et~al.\/}(2021)Saini, Zaleski \& Fuster]{saini2021direct}
{\sc \au{Saini, M}, \au{Zaleski, S} \& \au{Fuster, D}} \yr{2021} Direct
  numerical simulation of heterogeneous bubble nucleation.  \bt{In {\em 11th
  International Cavitation Symposium (CAV2021), 2021.\/}}.

\bibitem[Sinha {\em et~al.\/}(2022)Sinha, Narayan \&
  Srivastava]{sinha2022microlayer}
{\sc \au{Sinha, GK}, \au{Narayan, S} \& \au{Srivastava, A}} \yr{2022}
  \at{Microlayer dynamics during the growth process of a single vapour bubble
  under subcooled flow boiling conditions}.  \jt{Journal of Fluid Mechanics}
  \bvol{931},  \pg{A23}.

\bibitem[Strasberg(1959)]{strasberg1959}
{\sc \au{Strasberg, M}} \yr{1959}  \at{Onset of ultrasonic cavitation in tap
  water}.  \jt{The Journal of the Acoustical Society of America}
  \bvol{31}~(2),  \pg{163--176}.

\bibitem[Sullivan {\em et~al.\/}(2022)Sullivan, Dockar, Borg, Enright \&
  Pillai]{sullivan2022inertio}
{\sc \au{Sullivan, Patrick}, \au{Dockar, Duncan}, \au{Borg, Matthew~K},
  \au{Enright, Ryan} \& \au{Pillai, Rohit}} \yr{2022}  \at{Inertio-thermal
  vapour bubble growth}.  \jt{Journal of Fluid Mechanics}  \bvol{948},
  \pg{A55}.

\bibitem[Tryggvason {\em et~al.\/}(2011)Tryggvason, Scardovelli \&
  Zaleski]{tryggvason2011}
{\sc \au{Tryggvason, G}, \au{Scardovelli, R} \& \au{Zaleski, S}} \yr{2011} {\em
  Direct numerical simulations of gas--liquid multiphase flows\/}.
  \publ{Cambridge University Press}.

\bibitem[Urbano {\em et~al.\/}(2022)Urbano, Bibal \& Tanguy]{urbano2022semi}
{\sc \au{Urbano, Annafederica}, \au{Bibal, Marie} \& \au{Tanguy,
  S{\'e}bastien}} \yr{2022}  \at{A semi implicit compressible solver for
  two-phase flows of real fluids}.  \jt{Journal of Computational Physics}
  \bvol{456},  \pg{111034}.

\bibitem[Urbano {\em et~al.\/}(2018)Urbano, Tanguy, Huber \&
  Colin]{urbano2018direct}
{\sc \au{Urbano, A}, \au{Tanguy, S}, \au{Huber, G} \& \au{Colin, C}} \yr{2018}
  \at{Direct numerical simulation of nucleate boiling in micro-layer regime}.
  \jt{International Journal of Heat and Mass Transfer}  \bvol{123},
  \pg{1128--1137}.

\bibitem[Voinov(1976)]{voinov1976hydrodynamics}
{\sc \au{Voinov, OV}} \yr{1976}  \at{Hydrodynamics of wetting}.  \jt{Fluid
  dynamics}  \bvol{11}~(5),  \pg{714--721}.

\bibitem[Wilson(1982)]{wilson1982drag}
{\sc \au{Wilson, S}} \yr{1982}  \at{The drag-out problem in film coating
  theory}.  \jt{Journal of Engineering Mathematics}  \bvol{16}~(3),
  \pg{209--221}.

\bibitem[Zhang \& Nikolayev(2023)]{zhang2023time}
{\sc \au{Zhang, X} \& \au{Nikolayev, VS}} \yr{2023}  \at{Time-averaged approach
  to the dewetting problem at evaporation}.  \jt{Europhysics Letters}
  \bvol{142}~(3),  \pg{33002}.

\bibitem[Zou {\em et~al.\/}(2018)Zou, Gupta \& Maroo]{zou2018origin}
{\sc \au{Zou, A}, \au{Gupta, M} \& \au{Maroo, SC}} \yr{2018}  \at{Origin,
  evolution, and movement of microlayer in pool boiling}.  \jt{The journal of
  physical chemistry letters}  \bvol{9}~(14),  \pg{3863--3869}.

\end{thebibliography}

\end{document}